# A molecular state of correlated electrons in a quantum dot

Sokratis Kalliakos[1], Massimo Rontani[2*], Vittorio Pellegrini[1*], César Pascual García[1], Aron Pinczuk[3,4], Guido Goldoni[2], Elisa Molinari[2], Loren N. Pfeiffer[4] and Ken W. West[4]

[1]NEST INFM-CNR and Scuola Normale Superiore, 56126 Pisa, Italy

[2]S3 INFM-CNR and Department of Physics, University of Modena and Reggio Emilia, 41100 Modena, Italy

[3]Depts of Appl. Phys & Appl. Math. and of Physics, Columbia University, New York, New York 10027, USA

[4]Bell Laboratories, Alcatel-Lucent, Murray Hill, New Jersey 07974, USA

*e-mail: rontani@unimore.it; vp@sns.it

**Correlation among particles in finite quantum systems leads to complex behaviour and novel states of matter. One remarkable example is predicted to occur in a semiconductor quantum dot[1-3] (QD) where at vanishing density the Coulomb correlation among electrons rigidly fixes their relative position as that of the nuclei in a molecule[4-14]. In this limit, the neutral few-body excitations are roto-vibrations, which have either rigid-rotor or relative-motion character[15]. In the weak-correlation regime, on the contrary, the Coriolis force mixes rotational and vibrational motions. Here we report evidence of roto-vibrational modes of an electron molecular state at densities for which electron localization is not yet fully achieved. We probe these collective modes by inelastic light scattering[16-18] in QDs containing four electrons[19]. Spectra of low-lying excitations associated to changes of the relative-motion wave function -the analogues of the vibration modes of a conventional molecule- do not depend on the rotational state represented by the total angular momentum. Theoretical simulations via the configuration-interaction (CI) method[20] are in agreement with the observed roto-vibrational modes and indicate that such molecular excitations develop at the onset of short-range correlation.**

While vibrations in ordinary molecules consist in oscillations of the heavy nuclei around their equilibrium positions, electrons in QDs, such as those in Fig. 1, are described by a probability distribution function. In the absence of disorder, electrons become fully localized only in the limiting case of vanishing density where they form a rigid rotor. In the opposite non-interacting limit, the uncorrelated QD electrons experience significant Coriolis forces in the rotational states. We expect that at finite density, when the quantum-mechanical correlations are sufficiently strong[21], the relative motion gets decoupled from the rigid rotation of the system, yielding a sequence of molecular-like energy levels labelled by the vibrational ($v$) and angular momentum ($M$) quantum numbers, respectively, as shown in the right part of Fig. 2. The insensitivity of the energy of the vibrational modes to the value of $M$ thus provides the signature of the emergence of this correlated state in QDs at experimentally accessible density regimes.

To probe these molecular-like roto-vibrational modes of correlated electrons in QDs we have developed the experimental setup for inelastic light scattering as shown in Fig. 1. To improve the signal above the noise level, the experiments were performed in an array composed by $10^4$ nominally identical modulation-doped GaAs/AlGaAs QDs realized by nanolithography and dry etching. The homogeneity achieved in both lateral size and number of confined electrons of each QD of the array was demonstrated by micro-photoluminescence[22]. These nanostructures have an effective lateral size much smaller than their geometrical diameter $D$, due to the large depletion width of $\approx 90$ nm at the electron densities of $n = 1.1 \times 10^{11}$ cm$^{-2}$ of the modulation-doped quantum well used here[19,23]. In addition, the in-plane confinement potential can be well-approximated by a parabola with typical confinement energies in the range $\hbar\omega_0 = 1 \div 4$ meV (Ref. 19). This leads to a Fock-Darwin shell structure for non-interacting electrons[1], which has been observed by both transport[24] and inelastic light scattering experiments[16-18], and to the appearance of a Kohn mode in the far-infrared spectra[25].

In our experiment on QDs with four electrons (see Methods), the rotational state with angular momentum $M$ is tuned by application of a magnetic field[26,27]. In fact, according to Hund's rule, the four-electron ground state (GS) at $B = 0$ T is a spin triplet with zero total angular momentum, $(S, M)$ = (1, 0). The excitation spectrum in the spin channel (shown in the bottom part of Fig. 2) is therefore characterized by the inter-shell monopole spin excitation (peak $S_B$)[19], corresponding to $\Delta S$ = -1 and $\Delta M = 0$, from the $B = 0$ triplet GS to a singlet excited state (1,0)→(0,0). The application of a moderate magnetic field, $B$, perpendicular to the QD plane, induces a GS transition to a singlet state with $M = 2$, $(S, M) = (0, 2)$ (Ref. 24). This transition, expected when the cyclotron energy approximately equals the exchange term[24] (below $B \approx 0.5$ T for typical densities), appears in the collective spin spectra as a change of the signal at energies close to the $S_B$ peak (shaded area in Fig. 2). The increased scattering intensity at 0.2 - 0.3 T (inset to Fig. 2) is due to the emergence, around 5.5 meV, of three closely-spaced spin excitations of the new GS (0,2) as predicted by the CI calculations (arrows in the middle panel of Fig. 2, see Methods). The energy positions of all spin excitations are well reproduced by our model with $\hbar\omega_0 = 3.75$ meV, corresponding to the dimensionless density parameter (Coulomb-to-kinetic energy ratio) $r_S = 1.71$, where $r_S = 1/[a^*_B(\pi \cdot n)^{1/2}]$, $a^*_B$ is the effective Bohr radius, and the density $n$ is estimated as in Ref. 19. The GS spin transition is also in excellent agreement with the CI prediction of the transition taking place at $B = 0.276$ T (cf. the inset of Fig. 2). We remark that $B$ is here too small to enforce localization[26-28].

Signatures of formation of the roto-vibrational excitations of the correlated electron state, that are captured by the schematic energy level sequence reported to the right of Fig. 2, can be sought by comparing the excitations of the two GSs with $M = 0$ and $M = 2$, respectively. We focus on the low-lying spin and charge modes shown in Fig. 3 (left: experimental data; right: CI predictions). The key finding is that the lowest-energy spin excitation, i.e., $S_A$ for the (1,0) state and $S_C$ for the (0,2) state, does not shift as we go through the GS transition (panels of Fig. 3). As pointed out above, this is precisely the molecular signature in the QD, where the rigid rotation of the electrons is decoupled from the relative-motion dynamics. This experimental result is in sharp contrast with that theoretically expected in the absence of correlation. In fact, the $S_A$ and $S_C$ transitions occur at $\sim \hbar\omega_0$, and therefore are strongly renormalized with respect to the value of $\sim 2\hbar\omega_0$ at which the weak single-particle modes appear in the limit $r_S \rightarrow 0$. In addition, with no correlation, the $S_A$ and $S_C$ modes experience a large exchange-energy splitting $J \sim (\hbar\omega_0)^{1/2}$ (see Supplementary Information,

Fig. S2, Methods and Discussion), as confirmed by our self-consistent Hartree-Fock calculation[29] which predicts a splitting of ~1.5 meV for $r_S$ = 1.71 (red vertical lines in the left part of Fig. 3).

The low-lying charge mode for the (1,0) state ($C_A$ in Fig. 3) is replaced for the (0,2) state by a charge excitation ($C_B$) shifted at slightly higher energy while a replica of the spin mode $S_C$ appears in the charge channel due to the breakdown of the polarization selection rule induced by $B$ (see Supplementary Methods and Discussion and Figure S1). As for spin excitations (Fig. 2), the energies of charge excitations agree with those predicted by CI (Fig. 3 right) while the positions of $C_A$, $C_B$, and $S_B$ peaks are also in contrast with the HF calculations (not shown). The shortcomings of HF point to interpret the marginal energy shifts observed for the modes between the two GSs as a manifestation of strong correlation effects leading to the roto-vibrational structure of energy levels. Remarkably, as we argue below, this molecular signature occurs at $r_S$ values for which localization in space of electron wave functions is not yet fully achieved.

To gain insight into the relative motion of the electrons for the states experimentally accessed, we use CI to evaluate the pair correlation function $g(r)$ for the two GSs at theoretically extrapolated densities (Fig. 4a). The quantity $g(r)$ -the probability that two electrons are at distance $r$ (see the Methods section)- clearly shows that the internal motion of the electrons depends on $M$ when correlation is negligible ($r_s$ = 0.1, Fig. 4a), whereas it is substantially independent already at the experimental density of $r_s$ = 1.71. At very small densities ($r_s$ = 20, Fig. 4a) the two $g(r)$ values overlap completely and the molecule is a rigid rotor. This crossover is quantitatively studied by computing the functional distance separating $g(r)$ values of pairs of states, $\int_0^\infty dr |g_X(r) - g_{X'}(r)| r$, where $X$ ($X'$) is a state depending on $M = 0$ ($M = 2$). This is shown in Fig. 4b for the GSs (black curve) and the two lowest-energy spin excitations $S_A$, $S_C$ (red curve). The change in the slope of the functional distance, very close the experimental value of $r_s$ (the vertical bar in Fig. 4b), points to a transition from a liquid-like state at small $r_s$ to a molecule at large $r_s$. Remarkably, the critical value of $r_s$ is the same for both GSs and spin excitations.

The discovered transition to this correlated state is distinct from Wigner spatial localization of electrons, that emerges at larger values of $r_s$ in the intrinsic reference frame of the molecule. Wigner localization is seen by fixing the position of one electron in the *xy* plane (filled black circles in Fig. 4d), and then evaluating the conditional probability of measuring another electron[3] (see the Methods section). This conditional probability is plotted as contour plot in Fig. 4d for the *M* = 2 GS (left column) and the $S_C$ excited state (right column), respectively, for increasing values of $r_s$ (from top to bottom). Whereas in the non interacting case ($r_s = 0.1$) the only structure is the exchange hole around the fixed electron, at the experimental value of $r_s$ (centre panels of Fig. 4d) weight is moved away from the latter due to correlation. As $r_s$ is increased (bottom panels of Fig. 4d) electrons localize at the vertices of a square in the *M* = 2 GS, whereas the charge distribution of $S_C$ is consistent with the lowest-energy $B_1$ normal mode of vibration for the $C_{4v}$ point symmetry group of the square (white diagrams in Fig. 4d and Supplementary Information, Methods and Discussion).

To assess the threshold for Wigner localization, we also compute the spin-resolved probability density $n_\sigma(r)$ of the triplet GS with spin projection $S_z = 1$, and evaluate the functional distance between $n_+(r)$ and $n_-(r)$, $\int_0^\infty dr |n_+(r) - n_-(r)| r$, plotted in Fig. 4c (see the Methods section). As there are three spin-up and one spin-down electron, the difference is expected to vanish only in the limit $r_s \to \infty$, when the overlap among the wave functions of fully localized electrons becomes negligible as well as their mutual exchange interaction, making the spin degree of freedom irrelevant. In contrast with Fig. 4b, the variation of the slope of the curve in Fig. 4c is smooth with $r_s$, showing that no sharp boundary for electron localization can be found[14]. Exchange interaction between partially localized electrons also explains the fine structure of roto-vibrational levels highlighted in the energy scheme of Fig. 2. We have checked that for large values of $r_s$ the energy splitting between the excitations $S_A$, $S_B$, $S_C$, and $C_B$ becomes negligible and all the states collapse into the same $B_1$ roto-vibrational band (see Supplementary Information, Methods, Discussion and Table ST2).

Correspondence and requests for materials should be addressed to M. R. or V. P.

**ACKNOWLEDGEMENTS**

This work was supported by the projects MIUR-FIRB No. RBIN04EY74 and No. RBIN06JB4C, PRIN 2006 No. 2006022932 "Few-electron phenomena in semiconductor-quantum-dot devices", and the INFM-CINECA Supercomputing Project 2007. A. P. is supported by the National Science Foundation under Grant No. DMR-0352738, by the Department of Energy under Grant No. DE-AIO2-04ER46133, and by a research grant from the W. M. Keck Foundation. We thank F. Beltram, S. Corni, and R. Fazio for useful discussions.

**AUTHOR CONTRIBUTIONS**

The nanofabrication and experiments were carried out by S. K., V. P., C. P. G., and A. P.; M. R., G. G., and E. M., were responsible for the theoretical analysis; L. N. P. and K. W. W. provided the modulation-doped quantum well samples.

**METHODS**

**Experimental setup.** The sample was placed in a dilution magneto-cryostat reaching temperatures under illumination down to 200 mK. A tunable Ti:Sapphire continuous-wave single-mode laser with frequency $\omega_L$ (close to 1560 meV) was used as excitation source. The scattered light from the QD array at $\omega_S$ was collected through a series of optics, dispersed by a triple-grating spectrometer and detected by a charge-coupled device camera (cf. Fig. 1). Samples were fabricated from a 25 nm wide, one-side modulation-doped $Al_{0.1}Ga_{0.9}As/GaAs$ quantum well (density $n = 1.1 \times 10^{11}$ cm$^{-2}$ and mobility $\mu \approx 3 \times 10^6$ cm$^2$/Vs) by electron beam lithography and inductive coupled plasma reactive ion etching.

**Inelastic light scattering.** Neutral electronic excitations in GaAs QDs can be classified in terms of changes of total angular momentum $\Delta M$ and total spin ($\Delta S = 0$ for charge excitations, $\Delta S = \pm 1$ for spin excitations) and can be selectively probed by setting the linear polarizations (parallel for charge and perpendicular for spin) of the incident and scattered photons, respectively[17-19]. The parity selection rule dictates that monopole excitations with $\Delta M = 0$ are the strongest modes active in the inelastic-light-scattering experiments in the backscattering configuration. Partial breakdown of the polarization selection rule occurs in our QDs at finite values of an applied magnetic field (see

Supplementary Information, Methods and Discussion). By comparing spin and charge spectra, direct evaluation of the impact of few-body effects may be inferred by measuring their energy position and splitting[19].

**Evidence of the four-electron QD population.** In order to achieve the four-electron population, different QD arrays with *D* between 240 nm and 180 nm were nanofabricated. Identification of this number of electrons is linked to the observation of the spin mode labelled $S_B$ (cf. Fig. 2). The assignment of the $S_B$ mode is confirmed by the theoretical evaluations based on a full CI calculation[20] (cf. the Raman-active CI excitations labelled by the arrows in Fig. 2). Contrary to the other spin peaks seen in the spectrum of Fig. 2 at *B* = 0 T, the $S_B$ mode is not observed in other QDs with different *D*. The rapid disappearance of the $S_B$ mode above 2 K (not shown), due to the thermal occupation of a low-lying singlet state, independently confirms the nature of this peak[19].

**Calculation of the energy spectrum and wave functions.** We used the full CI approach[20] to solve with high numerical accuracy the few-body problem of *N* interacting electrons associated to the Hamiltonian:

$$H = \sum_{i=1}^{N} (\mathbf{p}_i - e\mathbf{A}(\mathbf{r}_i)/c)^2 / 2m^* + m^* \omega_0^2 (x_i^2 + y_i^2)/2 + V(z_i) + g_e \mu_B B s_{iz} + \sum_{i<j} e^2 / \kappa_r |\mathbf{r}_i - \mathbf{r}_j|.$$

Here the conduction band electrons are trapped in a dot confined in the *xy* plane by a harmonic potential with inter-level energy spacing $\hbar\omega_0$ = 3.75 meV as well as along *z* by the potential *V(z)* of a symmetric square quantum well (whose width is 25 nm and energy offset 250 meV). $\mathbf{r}_i \equiv (x_i, y_i, z_i)$ is the position of the *i*th electron, $\mathbf{p}_i$ its canonically conjugated momentum, $\mathbf{A}(\mathbf{r}) = B\hat{\mathbf{z}} \times \mathbf{r}/2$ is the vector potential giving rise to the magnetic field *B* along *z*, $m^*$ = 0.067 $m_e$ is the GaAs conduction band effective mass, $m_e$ and *e* are the free electron mass and charge, respectively, $g_e$ = -0.44 is the bulk GaAs gyromagnetic factor, $\mu_B$ is the Bohr magneton, $s_{iz}$ is the *z*-component of the spin of the *i*th particle, $\kappa_r = 12.4$ is the relative dielectric constant, and *c* is the speed of light. The eigenstates of *H* are superpositions of Slater determinants, $\prod_{i=1}^{N} \hat{c}^+_{\alpha_i} |0\rangle$, which are obtained by filling in the single-particle spin-orbitals α with the *N* electrons in all possible ways,

where the second-quantization operator $\hat{c}_\alpha^+$ creates an electron in level $\alpha \equiv (n, m, s_z)$ when applied to the vacuum, $|0\rangle$. Here $n$ and $m$ are the radial and azimuthal quantum numbers of Fock-Darwin orbitals[1], respectively, which we included up to the 10th energy shell. Such orbitals, multiplied by the ground state of the well $V(z)$, are the eigenstates of the non-interacting part of $H$. The whole interacting Hamiltonian $H$, a matrix with respect to the basis of the Slater determinants, is first block diagonalized, where the blocks are labelled by the total orbital angular momentum, $M$, total spin, $S$, and its $z$-projection, $S_z$ (the symmetry-breaking effect of the Zeeman term is neglected here). Finally, each block is diagonalized via Lanczos method[20], yielding both eigenvalues and eigenstates at low energy (the block maximum linear size was $5.1 \times 10^4$). The accuracy of the calculation was estimated by comparing the analytically known value of the dipole Kohn excitation mode, $\hbar\omega_0$ at $B = 0$ T, to that calculated via full CI. The relative error for excitation energies, in the worst case of $r_s = 22$, was less than $7 \times 10^{-3}$.

**Calculation of density and correlation functions.** The spin-resolved probability density $n_\sigma(r)$, where $\sigma = +,-$ and $r = \sqrt{x^2 + y^2}$, is computed as the quantum average over a given CI state, $n_\sigma(r) = \frac{1}{N_\sigma} \int dz \left\langle \sum_{i=1}^{N} \delta(\mathbf{r} - \mathbf{r}_i) \delta_{\sigma,\sigma_i} \right\rangle$, where $N_\sigma$ is the number of electrons with spin $\sigma$. The total density is $n(r) = [N_+ n_+(r) + N_- n_-(r)] / N$. The conditional probability plotted in Fig. 4d is defined as $P(x, y; x_0, y_0) = \frac{1}{N(N-1)} \left\langle \sum_{i,j=1}^{N} \delta(\mathbf{r} - \mathbf{r}_i) \delta(\mathbf{r}_0 - \mathbf{r}_j) \right\rangle$, where $(x_0, y_0)$ is fixed at the average value of $r$ and $z$, $z_0$ are fixed at the centre of the quantum well. The correlation function $g(r)$ is obtained by integration of $P$ over the center-of-mass coordinate, $g(r) = A \int P(x/2 + X, y/2 + Y; -x/2 + X, -y/2 + Y) dX dY$, where the pre-factor $A$ is chosen so that $\int_0^\infty dr g(r) r = 1$.

# Figures

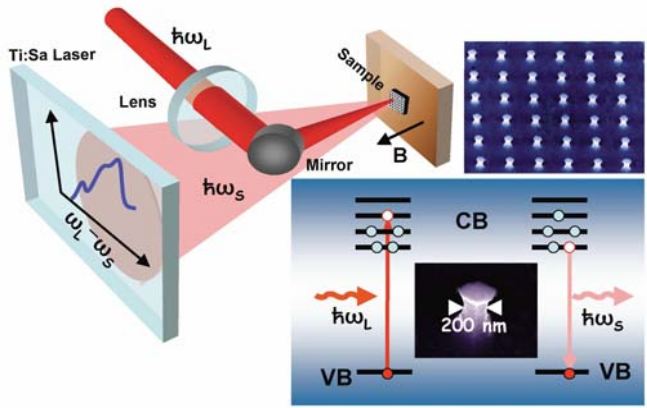

**Figure 1. Resonant inelastic light scattering.** (Top left panel) The inelastic light scattering set-up in the back-scattering geometry (see the Methods section). (Top right panel) SEM image of an array of QDs. (Bottom panel) A scheme of the experiment. An incident photon at $\hbar\omega_L$ resonating with a transition close to the gap between conduction (CB) and valence (VB) bands is absorbed creating an electron-hole pair and a scattered photon at $\hbar\omega_S$ is emitted with annihilation of a different electron-hole pair, leaving an excitation of energy $\hbar\omega_L - \hbar\omega_S$. The electronic configurations shown are the main CI contributions to the $S_C$ transition.

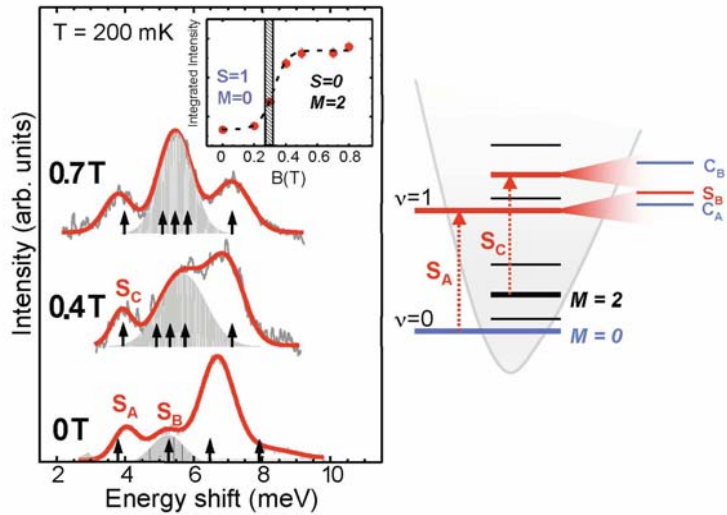

**Figure 2. Magnetic-field dependence in quantum dots with four electrons.** (Left panel) Experimental light scattering spectra of monopole spin excitations for three values of $B$. The red lines are fits to experimental data (grey lines) using three Gaussians (four at $B = 0$ T). The shaded areas correspond to a specific Gaussian. The arrows indicate the calculated excitations. The inset shows the integrated intensity of the central peak (shaded area) vs. $B$ (error bars correspond to standard deviation of the fits). The shaded box indicates the predicted position of the GS transition. (Right panel) A scheme of the roto-vibrational structure in the QD potential. Transitions $S_A$ and $S_C$ occur between different rotational states belonging to $\nu = 0$ and $\nu = 1$ vibrational levels (see Supplementary Information, Methods and Discussion).

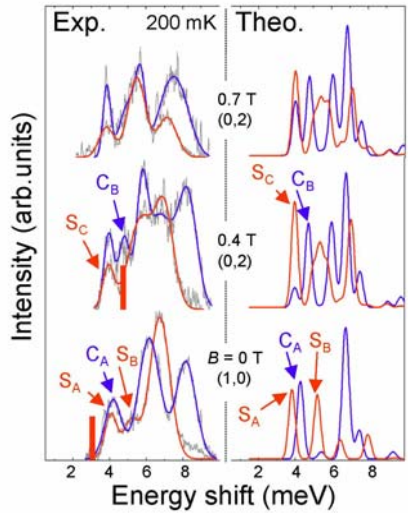

**Figure 3. Experimental / theoretical spectra with magnetic fields.** Experimental (left column) and computed (right column) spectra for charge (blue lines) and spin (red lines) excitations. The values (*S,M*) for the GSs are indicated. Blue and red lines in the left column are fits to the experimental data (gray lines) using Gaussians. The calculated peaks are artificially broadened using Gaussians with standard deviation 0.18 meV, and the laser energy used in the calculation, reckoned from the optical gap, is $\hbar\omega_L = 6$ (18) meV for charge (spin) excitations (see Supplementary Information, Methods, Discussion and Figure S1). Vertical red lines in the left panel are the Hartree-Fock predictions for $S_A$ and $S_C$.

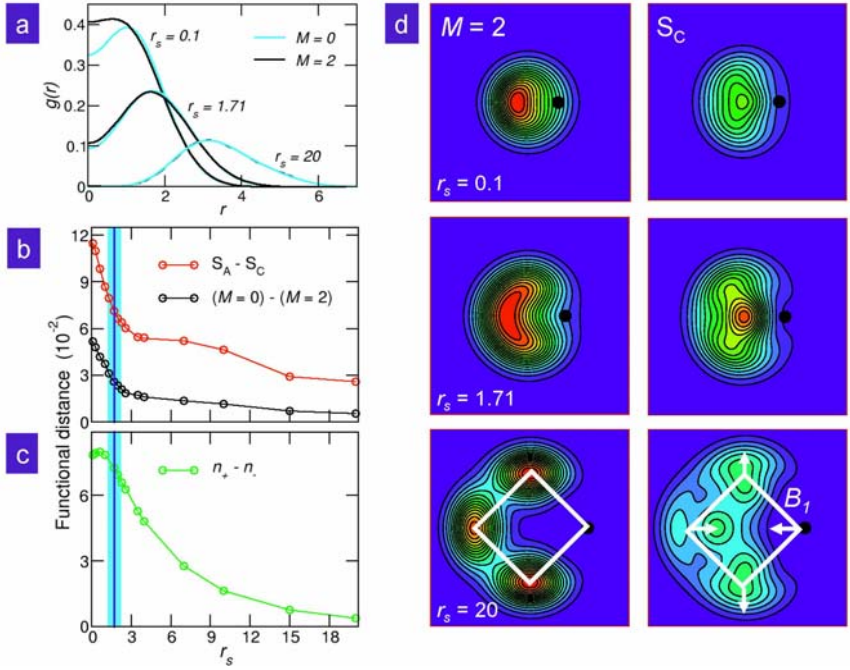

**Figure 4. Theoretical analysis of molecule formation extrapolated to the full density range.**
The unit length is $(\hbar/m^*\omega_0)^{1/2}$. **a,** $g(r)$ vs $r$ for the two GSs. **b,** Distance $\int_0^\infty dr |g_X(r) - g_{X'}(r)| r$ vs $r_s$ for two pairs of states $(X, X')$, where the first (second) pair consists in the two GSs ($S_A$ and $S_C$). The vertical bar width is obtained from the $B$-range of the GS transition. **c,** Distance between $n_+(r)$ and $n_-(r)$ vs $r_s$, for the triplet GS with $S_z = 1$. **d,** Probability of measuring an electron in the $xy$ plane provided another one is fixed at position $(x, y) = (x_0, 0)$ labelled by a black dot, where $x_0$ is located at the average value of $r$. The squares' size is $8 \times 8$, and the 15 equally-spaced contour levels go from blue (minimum) to red (maximum). The normalization is the same within each row.

# Supplementary Information

## List of contents

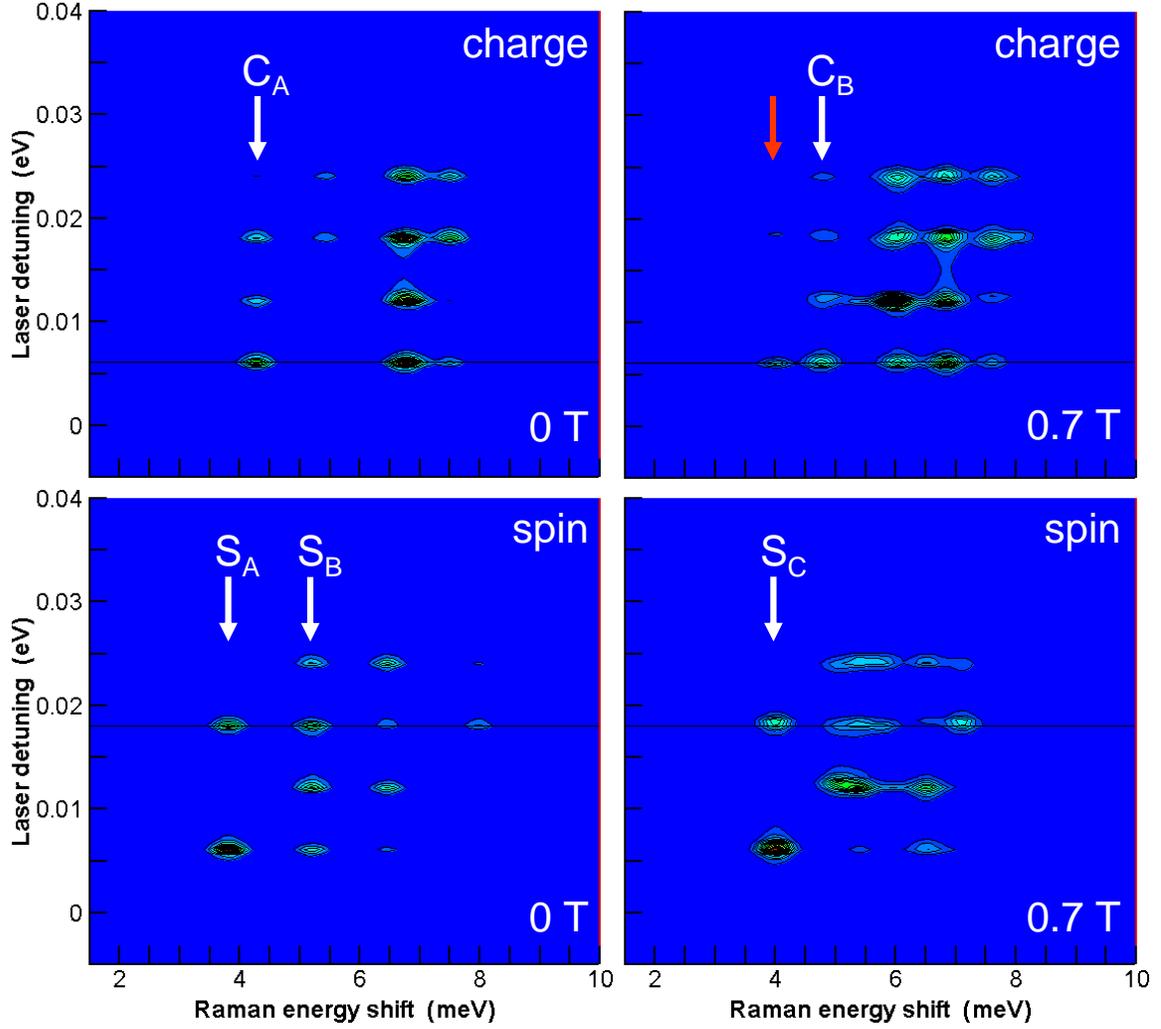

**Supplementary Figure S1. Calculated inelastic light scattering differential cross section as a function of both the measured energy shift and the incident laser energy.** Computed inelastic light scattering differential cross section, $d^2\sigma/d\Omega d\omega$, as a function of both the energy shift, $\hbar\omega$ (horizontal axis), and the incident laser energy reckoned from the optical gap, $\hbar\omega_L$ (vertical axis), for $B = 0$ (left column) and 0.7 T (right column), respectively. The contour level code goes from blue (minimum) up to green-yellow-red (maximum), the normalization of each plot is arbitrary, and the peaks were artificially broadened along the horizontal axis using Gaussian functions with standard deviation $\sigma = 0.18$ meV. The top (bottom) panels show excitations in the charge (spin) channel, and the strongest excitations are all monopolar. The horizontal black lines traced at $\hbar\omega_L = 6$ (18) meV in the top (bottom) plots correspond to the cuts along which the charge (spin) theoretical spectra of Fig. 3 were plotted. The excitations $C_A$, $C_B$, $S_A$, $S_B$, $S_C$, are highlighted (white arrows), together with the replica of the $S_C$ excitation in the charge channel (red arrow).
2

**Supplementary Methods and Discussion. Breakdown of polarization selection rules with the magnetic field.** The differential inelastic light scattering cross section, $d^2\sigma/d\Omega d\omega$, is proportional (at zero temperature) to a sum over excited states $F$, $d^2\sigma/d\Omega d\omega \propto \sum_F |M_{FI}|^2 \delta(\hbar\omega - E_F + E_I)$, where the matrix element $M_{FI}$ is the transition amplitude between the fully interacting ground and excited states $|I\rangle$ and $|F\rangle$, with energies $E_I$ and $E_F$, respectively, as obtained from the full CI calculation. $M_{FI}$ is defined as $M_{FI} = \sum_{\alpha\beta} \gamma_{\alpha\beta} \langle F|\hat{c}_\alpha^+ \hat{c}_\beta|I\rangle$, where $\gamma_{\alpha\beta}$ is the matrix element between α and β spin-orbitals within second order perturbation theory in the radiation field for the laser energy $\hbar\omega_L$ being close to the optical gap[19]:

$$\gamma_{\alpha\beta} \approx \frac{1}{m_e} \sum_{\beta'} \frac{\langle \alpha|\hat{\mathbf{e}}_L \cdot \mathbf{p} e^{i\mathbf{k}_L \cdot \mathbf{r}}|\beta'\rangle \langle \beta'|\hat{\mathbf{e}}_S \cdot \mathbf{p} e^{-i\mathbf{k}_S \cdot \mathbf{r}}|\beta\rangle}{\varepsilon_\alpha - \varepsilon_{\beta'} - \hbar\omega_L - i\Gamma_h}.$$

Here $\varepsilon_\alpha$ and $\varepsilon_{\beta'}$ are the energies of single-particle states $|\alpha\rangle$ and $|\beta'\rangle$ in the conduction and heavy hole valence band, respectively, including the Zeeman term coupling the spin with $B$, $\hat{\mathbf{e}}_L$ ($\hat{\mathbf{e}}_S$) and $\mathbf{k}_L$ ($\mathbf{k}_S$) are the polarization vectors and wave vectors of the incoming (scattered) photons, and $\Gamma_h$ is a phenomenological level width. $\gamma_{\alpha\beta}$ strongly enhances the cross section when the resonance condition is fulfilled, i.e., $\hbar\omega_L \approx \varepsilon_\alpha - \varepsilon_{\beta'}$. We took the same characteristic lateral (vertical) extension for both electron and hole Fock-Darwin (quantum well) states, $10^4$ cm$^{-1}$ for the in-plane component of the transferred momentum, $\mathbf{k}_S - \mathbf{k}_L$, and $\Gamma_h = 0.35$ meV.

At zero field, if the photon polarization vectors $\hat{\mathbf{e}}_L$ and $\hat{\mathbf{e}}_S$ are parallel (perpendicular), then $M_{FI}$ may differ from zero only if $\Delta S = 0$ ($|\Delta S| = 1$), where $\Delta S$ is the difference between the total spins of final and initial states, $|F\rangle$ and $|I\rangle$, respectively (see Methods and Ref. 19). If $B$ is finite, this selection rule breaks down [this has also been noticed in Delgado, A., and Gonzalez, A. & Lockwood, D. J. Selection and jump rules in electronic Raman scattering from GaAs/Al$_x$Ga$_{1-x}$As artificial atoms. *Phys. Rev. B* **71,** 241311(R) (2005)]. The reason is that the field lifts the degeneracy of the $J_z = \pm 1/2$ conduction band and $J = 3/2$, $J_z = \pm 3/2$ heavy hole band levels via the Zeeman energy term, where $J$ ($J_z$) is the (z-component of the) total angular momentum. This causes the effective operator $\sum_{\alpha\beta} \gamma_{\alpha\beta} \hat{c}_\alpha^+ \hat{c}_\beta$ coupling the states $|F\rangle$ and $|I\rangle$ to have no definite symmetry with respect to rotations in the spin space. The effect is non negligible only if: (i) The laser energy is at resonance with the optical band gap. (ii) The splitting $\Delta E_{Zeeman}$ between the excitation energies of the virtual electron-hole pairs created by the laser with different angular momenta (labeled σ⁺ and σ⁻, in the Faraday configuration, for $J_z$ = +1 and -1, respectively) is comparable or larger than the effective width of the hole energy levels, $\Gamma_h$:

$$\Delta E_{Zeeman} \approx \Gamma_h .$$

Here $\Delta E_{Zeeman}$ = |$g_e$ + $g_h$| $\mu_B$ $B$, $\mu_B$ is the Bohr magneton, $B$ is the field, $g_e$ and $g_h$ are the effective electron and hole gyromagnetic factors, respectively [cf. Eqs. (5) and (6) and the notations used in Snelling, M. J., and Blackwood, E., and McDonagh, C. J., and Harley, R. T. & Foxon, C. T. B. Exciton, heavy-hole, and electron g factors in type-I GaAs/Al$_x$Ga$_{1-x}$As quantum wells. *Phys. Rev. B* **45**, 3922-3925(R) (1992)]. For example, with GaAs bulk parameters at $B$ = 0.2 T, $\Delta E_{Zeeman}$ is 85 μeV. Since the value of $\Gamma_h$ is unknown, it is a sensitive free parameter of the theory: indeed, unrealistically small values of $\Gamma_h$ would artificially force the selection rule to break down at very small values of $B$. In the calculation, a reasonable value of $\Gamma_h$ = 0.35 meV was used, but actually the breakdown of selection rules occurs already at $B$ = 0.4 T even for values of $\Gamma_h$ as large as 1 meV.

In general, the calculated intensities of the spectra strongly depend on the resonance of the virtual electron-hole transitions with the laser energy, as it is clearly seen in Supplementary Figure S1, which displays the whole dependence of $d^2\sigma/d\Omega d\omega$ on laser energy $\hbar\omega_L$, for $B$ = 0 (left column) and 0.7 T (right column), respectively. The pictures are contour plots of the spectra intensities in a two-dimensional space, where the horizontal axis is the energy shift, $\hbar\omega$, and the vertical axis is the laser energy reckoned from the optical gap, $\hbar\omega_L$. The colour code for the equally spaced contour levels goes from blue (minimum) up to green-yellow-red, in order of increasing intensity, and the maximum height in each plot is different. The horizontal black lines traced at $\hbar\omega_L$ = 6 (18) meV in the top (bottom) plots correspond to the cuts along which the charge (spin) theoretical spectra of Fig. 3 were plotted. Remarkably, for the theoretical data of Fig. 3, the value $\hbar\omega_L$ of the laser energy for a given polarization, reckoned from the optical gap, is the same for all fields ($\hbar\omega_L$ = 6 meV for the charge channel, blue lines, and 18 meV for the spin channel, red lines in the right part of Fig. 3). From maps like Supplementary Figure S1, we are able to track down the excitations that are Raman active. In fact, the active excitations appear as vertical arrays of spots at a fixed peak energy position $\hbar\omega$, while the vertical inter-spot distance corresponds to the energy separation between different shells of levels from which electron-hole pairs are virtually created. Among the several arrays occurring in the plots, those corresponding to $C_A$ (top left panel), $S_A$, $S_B$ (bottom left), $C_B$ (top right), and $S_C$ (bottom right), are highlighted (white arrows). The left column of Supplementary Figure S1 displays maps for $B$ = 0 T, for which the ground state is the triplet and the polarization selection rule exactly holds. The right column shows the excitations of the singlet ground state at $B$ = 0.7 T. One clearly sees the replica of the $S_C$ signal in the charge channel (red arrow in the top right panel), which already appears at 0.4 T (data not shown). The major effect of the continuous variation of $B$ is that the replica of the $S_C$ peak in the charge channel increases its intensity with the field, while other changes due to the selection-rule break down are at first instance negligible.

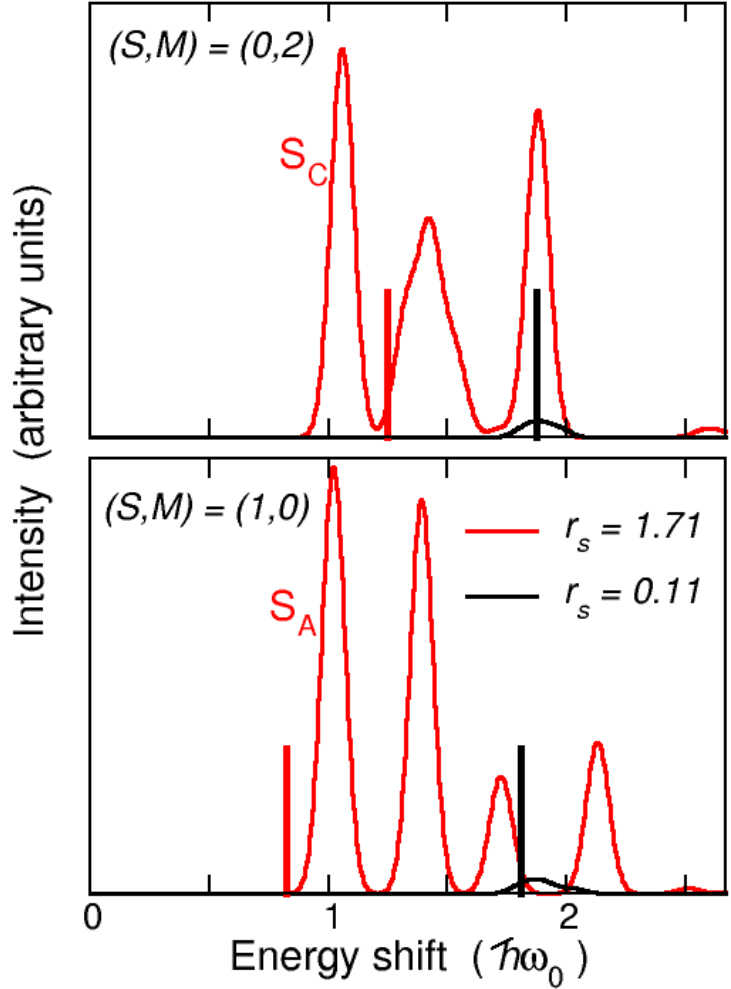

**Supplementary Figure S2. Spin excitations for $r_s = 1.71$ and $r_s = 0.11$: Configuration interaction vs Hartree-Fock predictions.** Computed inelastic light scattering differential cross section, $d^2\sigma/d\Omega d\omega$, as a function of the energy shift, $\hbar\omega$, for the spin excitations of the triplet and singlet ground states $(S,M) = (1,0)$ and $(0,2)$ (bottom and top panels, respectively). The solid curves (red for $r_s = 1.71$ and black for $r_s = 0.11$, respectively) are the CI predictions, while the vertical lines are the Hartree-Fock values for $S_A$ (bottom panel) and $S_C$ (top panel), computed self-consistently according to the method described in Ref. 29. Since $d^2\sigma/d\Omega d\omega$ depends on a few parameters (see Supplementary Method and Discussion. Breakdown of polarization selection rules with the magnetic field), in order to compare data at different $r_s$ as much consistently as possible, we changed the value of $\kappa_r$ by keeping fixed all the other parameters, and then mapped it to $r_s$. All the data were obtained at $B = 0$ T since diamagnetic effects are negligible across the triplet-to-singlet ground state transition.

**Supplementary Discussion. The regime of negligible correlation, $r_s \to 0$.** In the limit $r_s \to 0$ all the low-energy monopolar modes merge into the trivial non-interacting value of $2\hbar\omega_0$, as it may be seen in Supplementary Figure S2 for both CI and Hartree-Fock data for $r_s = 0.11$. In this regime, the relevant few-body states may be described to a good approximation as single Slater determinants. In particular, the configuration of the triplet GS $(S, M) = (1,0)$ is made of two electrons in the $(n,m) = (0,0)$ Fock-Darwin orbital, and the other two in the spin-orbitals $(0,1,+)$ and $(0,-1,+)$, respectively; the $(S,M) = (0,2)$ GS has two electrons in the $(0,0)$ level and the other two in the $(0,1)$ level; the $S_A$ final state is made of $(0,0,+)$, $(0,1,+)$, $(0,-1,+)$, $(1,0,+)$ spin-orbitals; the $S_C$ final state is made of $(0,0,+)$, $(0,1,+)$, $(0,1,-)$, $(1,0,+)$ spin-orbitals. It is straightforward to compute the Hartee-Fock energy separation between $S_A$ and $S_C$ excitations in this limit as a difference of the average values of $H$ on the pertinent Slater determinants. The resulting value, $J$, is the sum of two exchange integrals, $J = J_{(0,0),(0,1)} + J_{(0,1),(1,0)}$, where $J_{(n,m),(n',m')}$ is the Coulomb exchange integral[29] between the Fock-Darwin orbitals $(n,m)$ and $(n',m')$. The final result, in the two-dimensional case, is given by $J = 0.450\sqrt{2R^*\hbar\omega_0}$, where $R^* = e^4 m^*/(2\kappa_r^2\hbar^2)$ is the effective Rydberg. Therefore, in units of $\hbar\omega_0$, $J \sim (\hbar\omega_0)^{-1/2}$, which increases as $\hbar\omega_0$ decreases ($r_s$ increases). The increase of $J$ with $r_s$ is clearly seen in Supplementary Figure S2 by comparing the Hartree-Fock values for $r_s = 0.11$ and $1.71$, respectively. The two-dimensional formula would predict a value of $J \sim 0.8\ \hbar\omega_0$ at $r_s = 1.71$. However, the actual data displayed (red vertical lines) are taken from a fully self-consistent three-dimensional calculation[29], which provides a reduced value of $J \sim 0.43\ \hbar\omega_0$ due to the relaxation of the spin-orbitals. Remarkably, the energy shift $J$ is almost entirely compensated by the correlation energy, which is fully taken into account in the CI data of Supplementary Figure S2 (cf. the red curves for $r_s = 1.71$).

The second, remarkable effect of Coulomb correlation on the inelastic light scattering spectrum is that the intensity of the latter strongly depends on $r_s$. This is seen again in Supplementary Figure S2, where the CI intensity of the excitations for both ground states (bottom and top panels) are reduced by more than one order of magnitude as $r_s$ decreases from 1.71 to 0.11. This is another genuine effect of correlation, since the brilliancy of the spectrum under resonant condition is provided by the enhancement factor $\gamma_{\alpha\beta}$, which is linked to a specific matrix element between initial and final few-body states (cf. Supplementary Methods and Discussion. Breakdown of polarization selection rules with the magnetic field). It turns out that the relevant matrix element is a cross-term in the expansion of the CI initial and final states on the basis of Slater determinants. This cross-term goes to zero when the expansion of the two state is limited to a single Slater determinant each, as it occurs in the limit $r_s \to 0$ of no correlation.

**Supplementary Methods and Discussion. Identification of the roto-vibrational modes of the four-electron molecule for large values of $r_s$.** The quantum states of few electrons confined in a two-dimensional harmonic trap are generally not trivial since, in the total Hamiltonian, the kinetic energy operator, which scales like $\sim r_s^{-2}$, does not commute with the Coulomb energy operator, which scales like $\sim r_s^{-1}$. In the molecular regime, excited states are obtained from the quantization of either the rigid rotation of the electron molecule as a whole (quantum number $M = 0, \pm 1, \pm 2, \ldots$) or the internal motion of electrons in a set of relative coordinates (quantum numbers $v_\alpha = 0, 1, 2, \ldots$, where $\alpha$ labels the specific vibrational motion). In fact, rotational and vibrational motions are decoupled as far as Coriolis effects, which appear in the intrinsic reference frame of the molecule, can be discarded. Additionally, monopolar excitations, for which $\Delta M = 0$, are independent from $B$ provided that diamagnetic effects and Zeeman coupling may be neglected. Only in the limiting (Wigner) case of very large $r_s$, the few-body problem may be easily solved, since then the kinetic operator may be neglected in the Hamiltonian to a first approximation. In this Wigner regime, the four-electron GS maps onto the classical equilibrium configuration consisting in the electrons being localized at the vertices of a square, while the vibrations are the small oscillations of the electrons around their equilibrium positions [Schweigert, V. A. & Peeters, F. M. Spectral properties of classical two-dimensional clusters. *Phys. Rev. B* **51**, 7700-7713 (1995)]. See also Supplementary Figure S3, Tables ST1 and ST2.

The theoretical evolution of the roto-vibrational modes of the four-electron molecule as a function of $r_s$ was monitored by carefully checking the CI energy, the one-particle density $n(r)$, the pair correlation function $g(r)$, and the conditional probabilities of the lowest-energy states. For large values of $r_s$, it was possible to straightforwardly link the quantum states to the excitations of the Wigner molecule made of electrons localized at the vertices of a square (cf. bottom row of Fig. 4d). In this limit, we numerically checked that the CI excited state energies at zero field coincide with those predicted analytically, $E(M, v_\alpha)$, which are the sum of two contributions[12], i.e., the rotational energy of a rigid symmetric top, $\hbar^2 M^2 / 2I$, plus the oscillator energy of the normal modes of vibration, $\sum_\alpha \hbar \omega_\alpha v_\alpha$. Here $I$ is the momentum of inertia of the molecule, $\hbar \omega_\alpha$ is the energy quantum of the $\alpha$th normal mode of vibration, and $E(M, v_\alpha) = \hbar^2 M^2 / 2I + \sum_\alpha \hbar \omega_\alpha v_\alpha + E(0,0)$. A table of the analytically computed frequencies of the normal modes of vibration is provided in Supplementary Table ST1, while the corresponding electron oscillations are schematically depicted in Supplementary Figure S3.

The identification of the CI states computed at large $r_s$ with the roto-vibrational modes of the square Wigner molecule is also confirmed by the following group-theoretical analysis[12], which parallels the calculation of the statistical weights of polyatomic molecules [cf. Chap. 13 of Ref. 15]. We assume that the only relevant motion of the four-electron molecule, in the Wigner limit, is given by the small oscillations of the electrons around their equilibrium positions. Consequently, the total wave function is the product of three terms: the first one only depends on the rotational coordinates, the second one on the vibrational coordinates, and the third one on the spin coordinates, respectively.

Besides, the total wave function must comply with the $C_{4v}$ point symmetry of the square plus it must be antisymmetric for the exchange of two electrons. The above requirements cause the symmetry of the total wave function to be $B_2$ for spin quintuplets, $E$ or $A_2$ for triplets, $A_1$ or $B_2$ for singlets. On the other hand, the rotational states carry symmetry $A_1$ or $A_2$ for $M \equiv 0$, $E$ for $M \equiv 1,3$, $B_1$ or $B_2$ for $M \equiv 2$, where the congruent sign in these terms, $\equiv$, refers to modulo 4. Since all quantum states are labelled according to the values of $(S, M, v_\alpha)$, i.e., the total spin, orbital angular momentum, and number of excitation quanta of the αth vibrational mode, respectively, it is straightforward to determine, e.g., which values of $S$ are consistent with a given set of quantum numbers $(M, v_\alpha)$. Supplementary Table ST2 reports the allowed values of $S$ as a function of $(M, v_\alpha)$, with $\sum_\alpha v_\alpha \leq 1$. These symmetry constraints imply, e.g., that the states forming the first $B_1$ rotational band may only be either singlets or quintuplets for $M = 0$ and either singlets or triplets for $M = 2$. This necessary condition to belong to the $B_1$ band is satisfied for $S_A$, $S_B$, $S_C$, and $C_B$. Similar arguments allow to exclude that these same excitations may belong to a different rotational band.

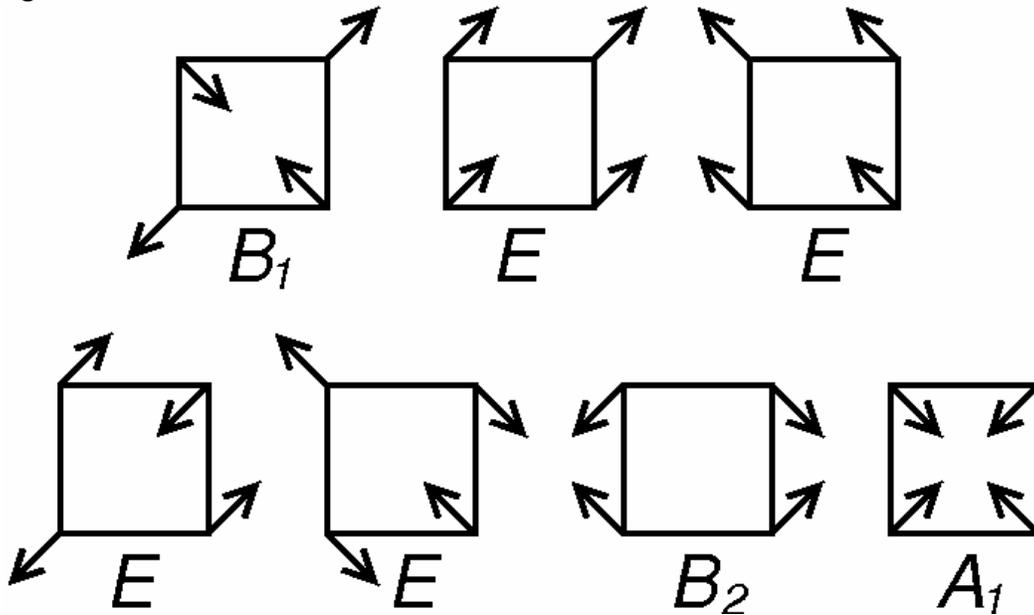

**Supplementary Figure S3. Normal modes of vibration of the four-electron Wigner molecule.** Schematic representation of the seven normal modes of vibration of four classical point-like particles in a two-dimensional harmonic trap interacting via repulsive Coulomb interaction, in order of increasing energy (from the left top to the right bottom). The eigth mode -at zero energy- is the rigid rotation of the molecule as a whole (not shown). The two-fold degenerate second and third $E$ modes are center-of-mass oscillations (Kohn mode), while the high-energy symmetric mode $A_1$ is the breathing mode. The oscillations are labelled according to the irreducible representations of the $C_{4V}$ point-symmetry group. The mode frequencies are reported in Table ST1.

| Mode | Multiplicity | Frequency ($\omega_0$ units) |
|---|---|---|
| $B_1$ | 1 | $\sqrt{\dfrac{3}{2\sqrt{2}+1}}$ |
| $E$ | 2 | 1 |
| $E$ | 2 | $\sqrt{\dfrac{4\sqrt{2}+1}{2\sqrt{2}+1}}$ |
| $B_2$ | 1 | $\sqrt{\dfrac{6\sqrt{2}}{2\sqrt{2}+1}}$ |
| $A_1$ | 1 | $\sqrt{3}$ |

**Supplementary Table ST1. Frequencies of the normal modes of vibration of the four-electron Wigner molecule.** The frequencies of the modes depicted in Supplementary Figure S3 are here reported, expressed in units of the confinement frequency of the two-dimensional harmonic trap, $\omega_0$.

| Vibrational mode excited | $M \equiv 0$ | $M \equiv 1,3$ | $M \equiv 2$ |
|---|---|---|---|
| No vibration | $S = 0, 1$ | $S = 1$ | $S = 0, 2$ |
| $B_1$ | $S = 0, 2$ | $S = 1$ | $S = 0, 1$ |
| $E$ | $S = 1$ | $S = 0, 1, 2$ | $S = 1$ |
| $B_2$ | $S = 0, 2$ | $S = 1$ | $S = 0, 1$ |
| $A_1$ | $S = 0, 1$ | $S = 1$ | $S = 0, 2$ |

**Supplementary Table ST2. Allowed quantum numbers for the low-energy excited states of the four-electron Wigner molecule.** The table shows the possible quantum numbers of the excited states of the four-electron Wigner molecules according to group theory (cf. Supplementary Methods and Discussion. Identification of the roto-vibrational modes of the four-electron molecule for large values of $r_s$). The states belong to either the "yrast" line (no quanta of vibration excited), or to the first-excited rotational bands (one single quantum of vibration excited). Here $M$ is the total orbital angular momentum, $S$ is the total spin, and the congruent sign, $\equiv$, refers to modulo 4.